\documentclass[12pt]{article}

\usepackage{amsmath,amsfonts,amssymb,latexsym}

\newtheorem{theorem}{Theorem}

\numberwithin{equation}{section}
\def\be{\begin{equation}}
\def\ee{\end{equation}}
\def\bq{\begin{eqnarray}}
\def\eq{\end{eqnarray}}
\def\beq{\begin{eqnarray*}}
\def\eeq{\end{eqnarray*}}
\def\f{\phi}
\def\c{\chi}
\def\a{\alpha}
\def\b{\beta}
\def\g{\gamma}
\def\d{\delta}
\def\l{\lambda}
\def\m{\mu}

\def\na{\nabla}
\def\pa{\partial}

\begin{document}
\begin{titlepage}
\begin{flushright}
\end{flushright}

\vspace{0.7cm}

\begin{center}
{\huge Slice Energy and Theories of Gravitation}

\vspace{1cm}

{\large Spiros Cotsakis}\\

\vspace{0.7cm}

{\normalsize {\em Research Group of Cosmology, Geometry and
Relativity}}\\ {\normalsize {\em Department of Information and
Communication Systems Engineering}}\\ {\normalsize {\em University
of the Aegean}}\\ {\normalsize {\em Karlovassi 83 200, Samos,
Greece}}\\ {\normalsize {\em E-mail:} \texttt{skot@aegean.gr}}
\end{center}

\vspace{0.7cm}

\begin{abstract}
\noindent We review recent work on the use of the slice energy
concept in generalized theories of gravitation. We focus on two
special features in these theories, namely, the energy exchange
between the matter component and the scalar field generated by the
conformal transformation to the Einstein frame of such theories
and the issue of the physical equivalence of different conformal
frame representations. We show that all such conformally-related,
generalized theories of gravitation allow for the slice energy to
be invariably defined and its fundamental properties be
insensitive to conformal transformations.
\end{abstract}

\vspace{1cm}
\begin{center}
{\line(5,0){280}}
\end{center}

\end{titlepage}

\section{Introduction}
Higher-order and scalar-tensor theories of gravity are currently
very popular as mechanisms providing alternative ways to explain
the observed cosmic acceleration, cf. \cite{cot04a,cot04b} and
Refs. therein. Although different these theories share two
important common characteristics: Firstly they can all be
formulated in different conformal frames and secondly they all
require for their proper formulation in at least one of the
conformally-related frames the existence of scalar fields. These
two characteristics turn out to be closely related: the conformal
transformation that relates two different conformal
representations of a theory is usually defined through the
introduction of a scalar field. Further the existence of different
conformal frames poses nontrivial relations between the scalar
fields present in them, which would otherwise have no connection.
It is therefore important to be able to state clearly such
relationships: What is the precise relation between the scalar
fields present in two different, conformally-related frames? Are
two frame representations of the same theory mathematically and/or
physically equivalent?

In this paper we first analyze these questions from the viewpoint
of a geometric quantity, the energy of fields on a slice in
spacetime (slice energy). In the next Section we review the basic
properties of this quantity and derive an important equation which
describes how the field energy is transported from slice to slice
in spacetime. In Section 3, we analyze the behaviour of the slice
energy in different theories of gravity and compare our findings
with that known in general relativity. This comparison shows that
slice energy is a kind of `universal invariant' in metric theories
of gravitation may be used to clarify the relations between the
different forms of scalar fields appearing in such theories.

The conformal technique further  means that we may view the
$f(R)$-vacuum theory as a unified theory of gravitation (described
in the Einstein frame by the metric $\tilde{g}$) and the scalar
field $\f$, that is as a theory uniting general relativity and the
(lagrangian) theory of the scalar field. In the Jordan frame only
one single  geometric object appears, the `metric' $g$, and the
conformal transformation then serves as a tool to `fragment' $g$
into its two pieces in the Einstein frame, namely the
gravitational field $\tilde{g}$ and the scalar field $\f$. In
higher-order gravity we may describe the different pieces of
information involved in the conformal transformation in the
following way: \emph{Gravity}, the field $\tilde g$  (the field
$g$ contains more than pure gravity); \emph{dark energy}, the
scalar field $\f$; \emph{dark matter}, the  ``matter'' fields
$\psi$ which couple non-minimally to gravity and to $\f$;
\emph{ordinary matter}, the conformal transform of the matter
component, $\tilde\psi$, which couples minimally to gravity but is
not coupled to $\f$. We show in the last Section how this
interaction, induced in the Einstein frame by adding, as an
example,  a dust cloud in the Jordan frame and conformally
transforming, leads to an exchange of energy between the this dust
cloud and the `unobserved' component--the $\f -$field and vice
versa. This in turn translates to an interplay between the matter
component in the original Jordan frame (dark matter) and the
scalar field $\f$ (dark energy).

\section{Slice energy and transport}
Consider a time-oriented spacetime $(\mathcal{V},g)$ with
$\mathcal{V}=\mathcal{M} \times \mathbb{R},$ where $\mathcal{M}$
is a smooth manifold of dimension $n$, $g$ a spacetime metric and
the spatial slices $\mathcal{M}_{t}\,(=\mathcal{M}\times \{t\})$
are spacelike submanifolds endowed with the time-dependent spatial
metric $g_{t}$. (In the following, Greek indices run from $0$ to
$n$, while Latin indices run from $1$ to $n$. We also assume that
the metric signature is $(+-\cdots -)$.) On $(\mathcal{V},g)$ we
consider a family of matterfields denoted collectively as $\psi $,
assume that the field $\psi$ arises from a lagrangian density
which we denote by $L$ and denote the stress tensor of the field
$\psi$  by $T(\psi )$.

For $X$ any causal vectorfield of $\mathcal{V}$ we define the
\emph{energy-momentum vector $P$ of a stress tensor $T$ relative
to $X$} to be \begin{equation}\label{p} P^{\b}=X_{\a}T^{\a\b}.
\end{equation} The \emph{energy on $\mathcal{M}_{t}$ with respect
to $X$}  is defined by the integral (when it exists)
\begin{equation}
E_{t}=\int_{\mathcal{M}_{t}}P^{\a}n_{\a}d\m_{t}, \end{equation}
where $n$ is the unit normal to $\mathcal{M}_{t}$ and $d\m_{t}$ is
the volume element with respect to the spatial metric $g_{t}$. We
call $P^{\a}n_{\a}$ the \emph{energy density}. Assuming that $X$
and $T$ are smooth, we find
\begin{equation}
\na_{\a}P^\a
=\frac{1}{2}T^{\a\b}(\na_{\a}X_{\b}+\na_{\b}X_{\a})+X_{\b}\na_{\a}T^{\a\b}.
\end{equation} Thus, if $\mathcal{K}\subset\mathcal{V}$ is a compact domain
with smooth boundary $\pa\mathcal{K}$, it follows from Stokes'
theorem that
\begin{equation}
\int_{\mathcal{K}}\na_{\a}P^\a d\mu
=\int_{\pa\mathcal{K}}P^{\a}n_{\a}d\sigma, \end{equation} where
$d\mu$ is the volume element of $\mathcal{V}$ and $d\sigma$ that
of $\pa\mathcal{K}$,
 and so we find
\begin{equation}
\int_{\pa\mathcal{K}}P^{\a}n_{\a}d\sigma=
\frac{1}{2}\int_{\mathcal{K}}T^{\a\b}(\na_{\a}X_{\b}+\na_{\b}X_{\a})d\mu+
\int_{\mathcal{K}}X_{\b}\na_{\a}T^{\a\b}d\mu. \end{equation}
Hence, when $\mathcal{M}$ is compact or the field falls off
appropriately at infinity, on the spacetime slab
$\mathcal{D}=\Sigma\times[t_{0},t_{1}]$,
$\Sigma\subset\mathcal{M}$, and with $T$ having support on
$\mathcal{D}$ we have the following relation for the energies on
the two end-slices \begin{equation}\label{trans-eq}
E_{t_{1}}-E_{t_{0}}=\frac{1}{2}\int_{t_{0}}^{t_{1}}\int_{\mathcal{M}_{t}}
T^{\a\b}(\na_{\a}X_{\b}+\na_{\b}X_{\a})d\mu
+\int_{t_{0}}^{t_{1}}\int_{\mathcal{M}_{t}}
X_{\b}\na_{\a}T^{\a\b}d\mu. \end{equation} Thus we have shown the
following result (cf. \cite{ch-mo00}, p. 87-88).
\begin{theorem}\label{1}
When $X$ is a Killing vectorfield and the field is conserved,
i.e., $\na_{\a}T^{\a\b}=0$, we have
\begin{equation}E_{t_{1}}=E_{t_{0}}. \end{equation}
\end{theorem}
This means that, when the energy-momentum tensor of a field is
conserved, the same is true for its slice energy relative to a
Killing vectorfield as a function of time.

Below, we pay particular attention to the case for which  the
field is a matter field $\psi$ interacting with a scalar field
$\phi$ with potential $V(\phi)$. We take the scalar field
lagrangian density to be
\begin{equation}
L=-\frac{1}{2}g^{\a\b}\pa_{\a}\f\pa_{\b}\f +V(\phi).
\end{equation}
Then the energy-momentum tensor of $\phi$ is
\begin{equation}\label{t1}
T^{\a\b}(\f)=\pa^\a\phi\pa^\b\phi-\frac{1}{2}g^{\a\b}(\pa^\l\phi\pa_\l\phi-2V(\phi)),
\end{equation}
and before we proceed further we note the following important
result (see \cite{cot04a} for a proof).
\begin{theorem}\label{2}
The energy density $P^\a n_\a$ of the scalar field $\phi$ with
potential $V(\phi)$ is positive when   $V(\phi)>0$.
\end{theorem}

As noted already, we shall be interested in the case of a matter
field $\psi$ interacting with a scalar field $\phi$ with potential
$V(\phi)$, especially in a conformally equivalent frame. The
following notation for conformally related quantities is used: Let
$g$ and $\tilde g$ be two conformal metrics, $\tilde g =\Omega^{2}
g$ on the manifold $\mathcal{V}$. This means that in two
\emph{orthonormal} moving frames, $\theta^\a$ and
$\tilde\theta^\a$, the two conformal metrics satisfy
\begin{equation}
\tilde g =\eta_{\a\b}\tilde\theta^\a\tilde\theta^\b,\quad
g=\eta_{\a\b}\theta^\a\theta^\b\quad\textrm{and}\quad\tilde\theta^\a=\Omega\,\theta^\a
,
\end{equation}
with $\eta_{\a\b}=\textrm{diag} (+,-\cdots -)$ being the flat
metric. Setting $\Omega^{2} =e^{\f}$ we see that
$\tilde\theta^\a=e^{\f/2}\theta^\a$ and obviously
$\tilde\theta_\a=e^{-\f/2}\theta_\a$. The same rules are true for
any 1-form or vectorfield  on $\mathcal{V}$. We take $\tilde{
T}^{\a\b}(\f )$ and $\tilde{T}^{\a\b}(\tilde{\psi})$ to denote the
stress tensors of the two  fields $\f$ and $\tilde{\psi}$ and
assume that their sum is conserved
\begin{equation}\label{stress-cons}
\tilde{\na}_\a\left(\tilde{ T}^{\a\b}(\f
)+\tilde{T}^{\a\b}(\tilde{\psi} )\right)=0,
\end{equation}
but the two components are \emph{not} conserved separately, that
is
\begin{equation}
\tilde{\na}_\a \tilde{T}^{\a\b}(\f )\neq 0,
\end{equation}
and
\begin{equation}\label{3.9}
\tilde{\na}_\a \tilde{T}^{\a\b}(\tilde{\psi} )\neq 0,
\end{equation}
unless the conservation equations $\na_\a T^{\a\b}(\psi)=0$ for
the field $\psi$ in the original frame are conformally invariant.
This implies that there must be a nontrivial $\phi -\tilde\psi$
interaction between the matter field $\tilde{\psi}$ and the
$\phi$-field and an associated \emph{exchange of energy} between
$\phi$ and $\tilde\psi$. Writing Eq. (\ref{trans-eq}) for the
scalar field $\f$ and substituting for the last term in the
right-hand-side from Eq. (\ref{stress-cons}) we arrive at the
\emph{general energy transport equation} in the conformally
related frame
\begin{equation}\label{en1dust}
E_{t_{1}} (\f )-E_{t_{0}} (\f
)=\frac{1}{2}\int_{t_{0}}^{t_{1}}\int_{\mathcal{M}_{t}}
\tilde{T}^{\a\b}(\f)(\tilde{\na}_{\a}\tilde{X}_{\b}+\tilde{\na}_{\b}\tilde{X}_{\a})d\tilde{\mu}
-\int_{t_{0}}^{t_{1}}\int_{\mathcal{M}_{t}}
\tilde{X}_{\b}\tilde{\na}_{\a}\tilde{T}^{\a\b}(\tilde{\psi}
)d\tilde{\mu},
\end{equation}
with $ d\tilde{\m}$ being the volume element of $\tilde{g}$.

\section{Field equations and slice energy conservation}
We are interested below in a comparison of the conservation
properties of slice energy of certain fields in general
relativity, higher-order gravity theories and scalar-tensor
theories of gravitation. In general relativity we take the field
equations to be of the form \be\label{gr}
G_{\a\b}=T_{\a\b}(\f)+T_{\a\b}(\psi ), \ee where $G_{\a\b}$ is the
Einstein tensor, $\f$ is a scalar field with stress tensor given
by Eq. (\ref{t1}), $T(\psi)$ represents the stress tensor of a
field $\psi$, and we assume the conservation identities $\na_\a
T^{\a\b}(\f )= 0$ and $\na_\a T^{\a\b}(\psi)= 0.$ In higher-order
gravity theories we consider the Jordan-frame equations
\be\label{hog1} L_{\a\b}\equiv
f'R_{\a\b}-\frac{1}{2}g_{\a\b}f-\na_\a\na_\b
f'+g_{\a\b}\,\square\,_{g}f'=T_{\a\b}(\psi), \ee which, because
$\na_a L^{\a\b}=0$, imply the conservation identities $\na_\a
T^{\a\b}(\psi)=0.$ The Einstein frame representation of this
theory is \be\label{hog2}
\tilde{G}_{\a\b}=T_{\a\b}(\f)+\tilde{T}_{\a\b}(\tilde{\psi} ), \ee
where $\f =\ln f'$ and $T_{\a\b}(\f)$ is of the form (\ref{t1})
with $V(\phi)=(1/2)(f')^{-2}(Rf'-f)$, cf. \cite{ba-co88}. Here we
have the situation introduced in the previous Section wherein the
whole tensor in the right-hand-side is conserved, \be\label{3.7}
\tilde{\na}_\a\left(\tilde{ T}^{\a\b}(\f
)+\tilde{T}^{\a\b}(\tilde{\psi} )\right)=0, \ee but the two
components are not conserved separately, that is
\be
\tilde{\na}_\a \tilde{T}^{\a\b}(\f )\neq 0, \quad \tilde{\na}_\a
\tilde{T}^{\a\b}(\tilde{\psi} )\neq 0. \ee The field $\f$
appearing both in general relativity and in (the Einstein frame
representation of) higher-order gravity theories is in certain
contexts responsible for the existence of an inflationary period.
For concreteness we call it \emph{the inflaton} and distinguish it
from a scalar field, say  $\xi$, that may appear directly in the
Jordan frame equations (\ref{hog1}) in addition to the matterfield
$\psi$.

Lastly we take the defining equations of our scalar-tensor theory
to be the Brans-Dicke (BD) ones, with $\c$ denoting the BD scalar
field (everything we do below is valid if, instead of the BD
theory assumed here only for brevity, we consider the most general
scalar-tensor action having couplings of the form $h(\c)$, where
$h$ is any differentiable function of the field $\c$),
\be\label{st} S_{\a\b}\equiv\c\,
G_{\a\b}=T_{\a\b}(\c)+T_{\a\b}(\psi ). \ee The novel feature of
this equation is the requirement that, if in accordance with the
equivalence principle we assume that
\be
\na_\a T^{\a\b}(\psi)= 0, \ee \emph{only}, then, because $\na_\a
G^{\a\b}=0$ we find
\be
\na_\a S^{\a\b}=\na_\a T^{\a\b}(\c). \ee Here $T^{\a\b}(\c)$ is
not given by (\ref{t1}) but by a different, more complicated,
expression (cf. \cite{we72}, pp. 159-60). For definiteness below
we call the field $\c$ \emph{the dilaton} to distinguish it from
the other scalar fields appearing in the $f(R)$  Eqs.
(\ref{hog1}), (\ref{hog2}) and in general relativity, Eq.
(\ref{gr}). Many  currently popular string theories appear as
special cases of the scalar-tensor equations.

Starting from the general transport equation, Eq. (\ref{en1dust}),
we derive relations showing the dependence of the total slice
energy of the system on the special features of each one of the
three theories given by Eqs. (\ref{gr}), (\ref{hog1}) and
(\ref{hog2}) and (\ref{st}). Writing Eq. (\ref{en1dust}) for the
scalar field $\f$ and substituting from Eqs. (\ref{gr}) and the
conservation identity  for the terms $T(\f)$ and $X\na\, T$
respectively, we find \bq\label{gr-hog}
E_{t_{1}}(\f)-E_{t_{0}}(\f)&=&\int_{t_{0}}^{t_{1}}\int_{\mathcal{M}_{t}}
[G^{\a\b}-T^{\a\b}(\psi)]\,\na_{(\a}X_{\b)}d\mu \nonumber\\&=&
\int_{t_{0}}^{t_{1}}\int_{\mathcal{M}_{t}}G^{\a\b}\na_{\a}X_{\b}d\mu
-\int_{t_{0}}^{t_{1}}\int_{\mathcal{M}_{t}}T^{\a\b}(\psi)\na_{\a}X_{\b}d\mu.
\eq Using Stokes' theorem, the last term is just \be\label{12}
\int_{t_{0}}^{t_{1}}\int_{\mathcal{M}_{t}}T^{\a\b}(\psi)\na_{\a}X_{\b}d\mu=
\int_{\mathcal{M}_{t_{1}}}P^{\a}n_{\a}d\mu_{t_{1}}-
\int_{\mathcal{M}_{t_{0}}}P^{\a}n_{\a}d\mu_{t_{0}}
=E_{t_{1}}(\psi)-E_{t_{0}}(\psi), \ee and so, setting
$E_{t}(\f+\psi)=E_{t}(\f)+E_{t}(\psi)$, we find that in general
relativity the total slice energy of a system comprised of the
field $\f$ and a matter field $\psi$ depends on the Einstein
tensor as follows: \be\label{gr-e1}
E_{t_{1}}(\f+\psi)-E_{t_{0}}(\f+\psi)=
\int_{t_{0}}^{t_{1}}\int_{\mathcal{M}_{t}}G^{\a\b}\na_{\a}X_{\b}d\mu.
\ee Further, since $\na_\a G^{\a\b}=0$, integrating by parts and
using Stokes theorem we have
\be
\int_{t_{0}}^{t_{1}}\int_{\mathcal{M}_{t}}G^{\a\b}\na_{\a}X_{\b}d\mu=
\int_{\mathcal{M}_{t_{1}}}G^{\a\b}X_\a N_\b d\mu_{t_{1}}-
\int_{\mathcal{M}_{t_{0}}}G^{\a\b}X_\a N_\b d\mu_{t_{0}}, \ee
where $N$ is the unit normal to the slices. Using this form we
have the following result.

\begin{theorem}
The total slice energy of the system comprised of the scalar field
$\f$ and a matterfield $\psi$ satisfying the Einstein equations
(\ref{gr}), is given by \be\label{gr-e2}
E_{t_{1}}(\f+\psi)-E_{t_{0}}(\f+\psi)=
\int_{\mathcal{M}_{t_{1}}}G^{\a\b}X_\a N_\b d\mu_{t_{1}}-
\int_{\mathcal{M}_{t_{0}}}G^{\a\b}X_\a N_\b d\mu_{t_{0}}. \ee In
particular, when  $X$ is a Killing field of the metric $g$, the
total slice energy of the system  is conserved.
\end{theorem}
The terms of the form $\int_{\mathcal{M}_{t}}G^{\a\b}X_\a N_\b
d\mu_{t}$ represent a gravitational \emph{flux} through the slice
$\mathcal{M}_{t}$. When $X$ is a Killing field, the right hand
side of Eq. (\ref{gr-e1}) is zero and we have an integral
conservation law given by the equality of the two terms in the
right hand side of Eq. (\ref{gr-e2}) and this agrees with the
corresponding result originally given in \cite{synge}, Chap. VI,
and proved there via a different route.

The situation in higher-order gravity is in fact, despite the
different conservation laws, similar. In the Einstein frame we
have \bq\label{gr-hog2}
E_{t_{1}}(\f)-E_{t_{0}}(\f)&=&\int_{t_{0}}^{t_{1}}\int_{\mathcal{M}_{t}}
[\tilde{G}^{\a\b}-\tilde{T}^{\a\b}(\tilde{\psi})]\,\tilde{\na}_{(\a}\tilde{X}_{\b)}d\tilde{\mu}
-\int_{t_{0}}^{t_{1}}\int_{\mathcal{M}_{t}}
\tilde{X}_{\b}\tilde{\na}_{\a}\tilde{T}^{\a\b}(\tilde{\psi})d\tilde{\mu}\nonumber\\&=&
\int_{t_{0}}^{t_{1}}\int_{\mathcal{M}_{t}}\tilde{G}^{\a\b}\tilde{\na}_{\a}\tilde{X}_{\b}
d\tilde{\mu}
-\int_{t_{0}}^{t_{1}}\int_{\mathcal{M}_{t}}\tilde{T}^{\a\b}(\tilde{\psi})
\tilde{\na}_{\a}\tilde{X}_{\b}d\tilde{\mu}\nonumber\\
&-&\int_{t_{0}}^{t_{1}}\int_{\mathcal{M}_{t}}\tilde{X}_{\b}\tilde{\na}_{\a}
\tilde{T}^{\a\b}(\tilde{\psi})d\tilde{\mu}. \eq Using Stokes'
theorem, the middle term is \bq
\int_{t_{0}}^{t_{1}}\int_{\mathcal{M}_{t}}\tilde{T}^{\a\b}(\tilde{\psi})
\tilde{\na}_{\a}\tilde{X}_{\b}d\tilde{\mu}&=&
\int_{\mathcal{M}_{t_{1}}}\tilde{P}^{\a}\tilde{n}_{\a}d\tilde{\mu}_{t_{1}}-
\int_{\mathcal{M}_{t_{0}}}\tilde{P}^{\a}\tilde{n}_{\a}d\tilde{\mu}_{t_{0}}\nonumber\\
&-&\int_{t_{0}}^{t_{1}}\int_{\mathcal{M}_{t}}\tilde{X}_{\b}\tilde{\na}_{\a}
\tilde{T}^{\a\b}(\tilde{\psi})d\tilde{\mu}\nonumber\\
&=&E_{t_{1}}(\tilde{\psi})-E_{t_{0}}(\tilde{\psi})
-\int_{t_{0}}^{t_{1}}\int_{\mathcal{M}_{t}}\tilde{X}_{\b}\tilde{\na}_{\a}
\tilde{T}^{\a\b}(\tilde{\psi})d\tilde{\mu} \eq and so, setting
$E_{t}(\f+\tilde{\psi})=E_{t}(\f)+E_{t}(\tilde{\psi})$, we find
that in higher-order gravity, because of the marvelous fact that
the terms of the general form $\int \tilde{X}\tilde{\na}
\tilde{T}(\tilde{\psi})$ which were absent in general relativity
now precisely cancel each other, the total slice energy of a
system composed of the field $\f$ and a matter field
$\tilde{\psi}$ in the Einstein frame depends on the Einstein
tensor in the same way as before: \be\label{hog-e1}
E_{t_{1}}(\f+\tilde{\psi})-E_{t_{0}}(\f+\tilde{\psi})=
\int_{t_{0}}^{t_{1}}\int_{\mathcal{M}_{t}}\tilde{G}^{\a\b}\tilde{\na}_{\a}\tilde{X}_{\b}
d\tilde{\mu}. \ee Hence we arrive at  the following result.

\begin{theorem}
The total slice energy of the system composed of the scalar field
$\f$ and a matterfield $\tilde{\psi}$ satisfying the Einstein
equations (\ref{hog2}), is given by \be\label{hog-e2}
E_{t_{1}}(\f+\tilde{\psi})-E_{t_{0}}(\f+\tilde{\psi})=
\int_{\mathcal{M}_{t_{1}}}\tilde{G}^{\a\b}\tilde{X}_\a
\tilde{N}_\b d\tilde{\mu}_{t_{1}}-
\int_{\mathcal{M}_{t_{0}}}\tilde{G}^{\a\b}\tilde{X}_\a
\tilde{N}_\b d\tilde{\mu}_{t_{0}}. \ee In particular, when  $X$ is
a Killing field of the metric $\tilde{g}$, the total slice energy
of the system  is conserved.
\end{theorem}
Note that if we have a scalar field $\xi$ in addition to the
matter field $\psi$ present in the original Jordan frame of the
higher order gravity theory, then we obtain a  result similar to
that in general relativity but with $L_{\a\b}$ in place of the
Einstein tensor, namely, \be\label{hog-e3}
E_{t_{1}}(\xi+\psi)-E_{t_{0}}(\xi+\psi)=
\int_{\mathcal{M}_{t_{1}}}L^{\a\b} X_\a N_\b d\mu_{t_{1}}-
\int_{\mathcal{M}_{t_{0}}}L^{\a\b} X_\a N_\b d\mu_{t_{0}}. \ee
Then  terms of the form $\int_{\mathcal{M}_{t}}L^{\a\b}X_\a N_\b
d\mu_{t}$  represent a \emph{higher-order gravitational flux}
through the slice $\mathcal{M}_{t}$. When $X$ is a Killing field,
we again have an integral conservation law as before.

We now move to the analysis of the scalar-tensor theory
(\ref{st}).  In this case Eq. (\ref{en1dust}) becomes
\be\label{st-e1} E_{t_{1}}(\c )-E_{t_{0}}(\c
)=\int_{t_{0}}^{t_{1}}\int_{\mathcal{M}_{t}}
X_{\b}\na_{\a}S^{\a\b}d\mu+\int_{t_{0}}^{t_{1}}\int_{\mathcal{M}_{t}}
T^{\a\b}(\c)\na_{(\a}X_{\b)}d\mu, \ee and after some algebra we
find that
\be
E_{t_{1}}(\c )-E_{t_{0}}(\c
)=\int_{\mathcal{M}_{t_{1}}}S^{\a\b}X_{\b}N_{\a}d\mu_{t_{1}}
-\int_{\mathcal{M}_{t_{0}}}S^{\a\b}X_{\b}N_{\a}d\mu_{t_{0}}-
\int_{t_{0}}^{t_{1}}\int_{\mathcal{M}_{t}}
T^{\a\b}(\psi)\na_{(\a}X_{\b )}. \ee Since $\na_\a
S^{\a\b}=G^{\a\b}\na_{\a}\f$, the first two terms can be expressed
more simply as follows
\be
\int_{\mathcal{M}_{t_{1}}}S^{\a\b}X_{\b}N_{\a}d\mu_{t_{1}}
-\int_{\mathcal{M}_{t_{0}}}S^{\a\b}X_{\b}N_{\a}d\mu_{t_{0}}=
\int_{t_{0}}^{t_{1}}\int_{\mathcal{M}_{t}}S^{\a\b}\na_{\a}X_{\b}+
\int_{t_{0}}^{t_{1}}\int_{\mathcal{M}_{t}}G^{\a\b}X_{\b}\pa_{\a}\chi
. \ee Therefore using Eq. (\ref{12}) we find that
\beq\label{st-e3}
E_{t_{1}}(\chi+\psi)-E_{t_{0}}(\chi+\psi)&=&\int_{t_{0}}^{t_{1}}\int_{\mathcal{M}_{t}}
\chi G^{\a\b}\na_\a
X_\b+\int_{t_{0}}^{t_{1}}\int_{\mathcal{M}_{t}}G^{\a\b}X_\b\pa_\a\chi
\nonumber\\
&=&\int_{t_{0}}^{t_{1}}\int_{\mathcal{M}_{t}}G^{\a\b}(\chi\na_\a
X_\b +X_\b\pa_\a\chi )\nonumber\\
&=&\int_{t_{0}}^{t_{1}}\int_{\mathcal{M}_{t}}G^{\a\b}\na_\a (\chi
X_\b ) \eeq and we are led to the following result.
\begin{theorem}
The total slice energy of the dilaton-matter system  satisfying
the scalar-tensor equations (\ref{st}) is given by
\be\label{st-e2} E_{t_{1}}(\chi+\psi)-E_{t_{0}}(\chi+\psi)=
\int_{\mathcal{M}_{t_{1}}}S^{\a\b} X_\a N_\b d\mu_{t_{1}}-
\int_{\mathcal{M}_{t_{0}}}S^{\a\b} X_\a N_\b d\mu_{t_{0}}. \ee
\end{theorem}

In conclusion we have found the different forms that slice energy
takes in various classes of generalized theories of gravitation
which include higher-order gravity theories and scalar-tensor
ones. These forms may be described symbolically as follows:
\be\label{last} E_{t_{1}}(\lambda+\psi)-E_{t_{0}}(\lambda+\psi)=
\int_{\mathcal{M}_{t_{1}}}\Lambda^{\a\b}X_\a N_\b d\mu_{t_{1}}-
\int_{\mathcal{M}_{t_{0}}}\Lambda^{\a\b} X_\a N_\b d\mu_{t_{0}}.
\ee Here $\lambda$ denotes either an inflaton field, $\f$, which
couples to matter in general relativity or in  the Einstein frame
in higher-order gravity,  the scalar field $\xi$ which may appear
in the Jordan frame of higher-order gravity, or the dilaton $\chi$
in scalar-tensor theory, while $\Lambda$ is a gravitational
operator defining the left-hand-sides of the associated field
equations, that is, $\Lambda$ is $G^{\a\b}, \tilde{G}^{\a\b},
L^{\a\b}$ or $S^{\a\b}$ respectively. As a last remark we note
that one might think that since all these theories have equations
of motion of the form
\be
\Lambda_{\a\b}=T^{\textrm{total}}_{\a\b}, \ee and since
\be
E_{\textrm{total}}=\int_{\mathcal{M}_{t}}d\mu_{t}X^{\a}T_{\a\b}n^{\b},
\ee then \be\label{after}
E_{\textrm{total}}=\int_{\mathcal{M}_{t}}d\mu_{t}X^{\a}\Lambda_{\a\b}n^{\b},
\ee and so Eq. (\ref{last}) follows trivially from Eq.
(\ref{after}). What is wrong with this argument? The basic point
here is that it is not obvious that the slice energy will
necessarily satisfy a linear law of the form
$E_{\textrm{total}}=E(\phi)+E(\psi)$. In fact, that this is the
\emph{only} (currently feasible) way to do it follows from Eq.
(\ref{gr-hog}) using Eq. (\ref{12})  as we acknowledge it
immediately after that equation. Usually one has only one field in
general relativity and so the equivalent of Eq. (\ref{gr-e1}) is
obvious. To say that Eq. (\ref{gr-e1}) is valid when more than one
fields are present, one has to \emph{assume} that slice energy
will satisfy the above linear law. So the results in this Section
clear up this point and prepare the necessary ground for the more
complicated case of interacting (i.e., directly coupled) fields.

A closely related issue is the crucial fact that,  in the
different theories we consider, not only one typically has a
number of fields in the RHSs of the equations of motion but each
of these fields is \emph{not} separately conserved.  This gives an
added complication to the effect that in the equations involving
the slice energy one has the slice energy of one field appearing
in the LHS and the matter tensor of \emph{another} field on the
RHS, and  there is no conservation law in the RHS as that field is
not separately conserved. Of course one may \emph{guess} that the
same behavior as that previously found for general relativity will
be valid here. But then this is clearly not a proof. The analysis
in this Section clears up this situation completely in the more
general case of higher order and scalar tensor theories. The field
equations do not say that slice energy is an additive quantity,
only assume it. This is an important point clarified in this
analysis.

\section{Dust clouds in higher order gravity}
We now study th interaction and associated energy exchange between
$\phi$ and $\tilde{\psi}$ more closely by assuming that $\psi$ is
the simplest form of a fluid, a dust cloud on $(\mathcal{V},g)$
with 4-velocity $V_\a$ and stress tensor
\be
T_{\a\b,\,\textrm{dust}}=\rho V_\a V_\b, \ee satisfying the
$f(R)$\emph{-dust equations} in the Jordan frame, namely
\be\label{f-dust} f'R_{\a\b}-\frac{1}{2}g_{\a\b}f-\na_\a\na_\b
f'+g_{\a\b}\,\square\,_{g}f'=\rho V_\a V_\b, \ee with
\be\label{cons dust} \na_\a (\rho V^\a V^\b)=0. \ee After the
conformal transformation  we find \be\label{ein-frame-dust}
\tilde{G}_{\a\b}=\tilde{T}_{\a\b}(\f)+\tilde{\rho}\, \tilde{V}_\a
\tilde{V}_\b, \ee with  $\tilde{T}_{\a\b}(\f)$ given by (\ref{t1})
with tildes where appropriate, and
\be
\tilde{V_{a}}=e^{-\f/2} V_{a},\quad\tilde{\rho}=e^{s\f/2}\rho
,\quad s\in\mathbb{R}. \ee (We have set $\tilde\rho =\Omega^s\rho$
and since $\Omega^2=e^\f$, $\Omega^s =e^{s\f/2}$.) What is the
field equation satisfied by the scalar field $\f$? From Eq.
(\ref{ein-frame-dust}) the divergence of the stress tensor of $\f$
is minus that of the dust, but
\be
\tilde{\na}_\a (\tilde{\rho}\, \tilde{V}^\a \tilde{V}^\b )= \na_\a
(\tilde{\rho}\, \tilde{V}^\a \tilde{V}^\b )+A^\a_{\a\g}
\tilde{\rho}\, \tilde{V}^\g \tilde{V}^\b+A^\b_{\a\g}
\tilde{\rho}\, \tilde{V}^\a \tilde{V}^\g, \ee
 where
 \be
A^\a_{\b\g}=\frac{1}{2}\left(\delta^{\a}_{\b}\pa_{\g}\f
+\delta^{\a}_{\g}\pa_{\b}\f -g_{\b\g}g^{\a\d}\pa_{\d}\f\right).
 \ee
From these equations and Eq. (\ref{cons dust}) we deduce the
modified scalar field equation in the form \be\label{sc field eqn}
\pa^{\b}\f(\widetilde{\square}\f+V')+\frac{s+5}{2} \tilde{\rho}\,
\tilde{V}^\a \tilde{V}^\b \pa_\a\f
-\frac{1}{2}\tilde\rho\pa^\b\f=0. \ee Another way to derive  the
scalar field equation is as follows. Since \be\label{3.16}
\tilde{\na}_\a \tilde{T}^{\a\b}_{\textrm{dust}}=\tilde{\na}_\a
(\tilde{\rho}\, \tilde{V}^\a \tilde{V}^\b )=
\tilde{V}^\b\tilde{\na}_\a (\tilde{\rho}\, \tilde{V}^\a
)+\tilde{\rho}(\tilde{\na}_\a \tilde{V}^\b )\tilde{V}^\a , \ee
multiplying (\ref{3.16}) by $\tilde{V}_\b$ we have
\be
\tilde{V}_\b\tilde{\na}_\a
\tilde{T}^{\a\b}_{\textrm{dust}}=\tilde{\na}_\a (\tilde{\rho}\,
\tilde{V}^\a )\tilde{V}_\b \tilde{V}^\b +\tilde{\rho}
(\tilde{\na}_\a \tilde{V}^\b )\tilde{V}^\a \tilde{V}_\b \ee and,
since $\tilde{V}_\b \tilde{V}^\b =1$, we obtain
\be\label{pre-energy} \tilde{\na}_\a
\tilde{T}^{\a\b}_{\textrm{dust}} =\tilde{V}^\b\tilde{\na}_\a
(\tilde{\rho} \tilde{V}^\a )+\tilde{\rho} (\tilde{\na}_\a
\tilde{V}^\b )\tilde{V}^\a. \ee After some algebra we arrive at
the equation of motion for the scalar field $\f$ in the Einstein
frame, namely \be\label{sc field eqn1}
\pa^{\b}\f(\widetilde{\square}\f+V')+
\tilde{V}^{\b}\,\tilde{\na}_{\a}(\tilde{\rho}\,\tilde{V}^{\a})
+\tilde{\rho}\tilde{V}^\a (\tilde{\na}_\a \tilde{V}^\b )=0. \ee
Recalling that dust matter follows geodesics on the original
Jordan frame, $V^\a\na_\a V^\b=0$, we find that the last two terms
in this equation equal to the last two terms in Eq. (\ref{sc field
eqn}) and so we conclude that Eq. (\ref{sc field eqn1}) provides
an equivalent form of Eq. (\ref{sc field eqn}).

 Note that in the very special case where
\be\label{3.18} \tilde{V}_{\b}=\pa_{\b}\f, \ee which implies some
sort of `alignment' between the dust and the scalar field,  the
scalar field equation (\ref{sc field eqn}) becomes
\be
\widetilde{\square}\f+V'+\frac{s+4}{2}\tilde\rho =0. \ee This is
now easily compared to the more commonly used scalar field
equation, but we have to bear in mind that it has been obtained
from (\ref{sc field eqn}) by imposing the serious restriction
(\ref{3.18}).

We now study the behaviour of the total slice energy of the system
comprised of $\f$ and the dust component. We prove that the total
slice energy is conserved only when spacetime is stationary. We
choose $V=X$ so that \be\label{3.11} P^\a n_\a =X_\b n_\a\rho V^\a
V^\b=\rho V^\a n_\a. \ee Hence, from Eq. (\ref{pre-energy}),
applying Stokes' theorem we obtain \be\label{stokes}
\int_{\mathcal{K}}\tilde{\na}_\a (\tilde{\rho} \tilde{V}^\a
)d\tilde\mu =\int_{\pa\mathcal{K}}\tilde{\rho} \tilde{V}^\a
\tilde{n}_\a d\tilde\sigma , \ee Therefore Eq. (\ref{en1dust}),
after some manipulation becomes \be\label{3.26A} E_t (\f ) +E_t
(\mathrm{dust}) =E_0 (\f ) +E_0 (\mathrm{dust})
+\frac{1}{2}\int_{t_{0}}^{t_{1}}\int_{\mathcal{M}_{t}}
\tilde{T}^{\a\b}(\f )(\tilde{\na}_{\a}\tilde{V}_{\b}+
\tilde{\na}_{\b}\tilde{V}_{\a})d\tilde{\mu}, \ee where by
definition and Eq. (\ref{3.11}), for any $t$,
\be
E_t (\mathrm{dust})=\int_{\mathcal{M}_{t}}\rho V^\a n_\a d\m_{t}.
\ee We see that the last term in Eq. (\ref{3.26A}) can be zero
only when $V$ is a Killing vectorfield.
We therefore arrive at the following result about the total slice
energy with respect to the fluid itself.
\begin{theorem}
The total  slice energy with respect to the timelike vectorfield
$\tilde V$, tangent to the dust timelines, of the scalar
field-dust system satisfying the field equations (\ref{f-dust}),
satisfies \be\label{3.26AA} E_t (\f +\mathrm{dust}) =E_0 (\f +
\mathrm{dust})
+\frac{1}{2}\int_{t_{0}}^{t_{1}}\int_{\mathcal{M}_{t}}
\tilde{T}^{\a\b}(\f )(\tilde{\na}_{\a}\tilde{V}_{\b}+
\tilde{\na}_{\b}\tilde{V}_{\a})d\tilde{\mu}. \ee In particular the
slice energy of the scalar field-dust system is conserved when
$\tilde V$ is a Killing vectorfield of $\tilde g$.
\end{theorem}
We also conclude that the property of the conservation of slice
energy for dust is a conformal invariant. However, when $V$ is not
a Killing vectorfield, we see that there is a nontrivial
contribution to the slice energy coming from the stress tensor of
the scalar field generated by the conformal transformation. Note
that this contribution is also nonzero even in the special case
that Eq. (\ref{3.18}) is assumed for in that case the first term
in $\tilde{T}^{\alpha \beta }(\phi ) (\tilde{\nabla}_{\alpha }
\tilde{V}_{\beta }+\tilde{\nabla}_{\beta }\tilde{V}_{\alpha})$ is
zero because $\tilde{V}^{\alpha }\tilde{V}^{\beta }(\tilde{\nabla}_{\alpha }%
\tilde{V}_{\beta }+\tilde{\nabla}_{\beta}\tilde{V}_{\alpha})=0$,
but the whole combination is still not zero as there are
additional terms coming from the contributions of the other terms
in Eq. (\ref{t1}) (unless the fluid satisfies an extra condition
-- see \cite{cot04a}). This analysis can be extended to the more
general case of a perfect fluid interacting with a scalar field,
cf. \cite{cot04a}.

\section*{Acknowledgements} This work was supported by the joint Greek Ministry of
Education/European Union Research grant `Pythagoras' No. 1351 and
this support is gratefully acknowledged.

\end{document}